\documentclass[letterpaper,english,reprint, aps]{revtex4-1}
\usepackage[T1]{fontenc}
\usepackage[latin9]{inputenc}
\usepackage{color}
\usepackage{amsmath}
\usepackage{amssymb}
\usepackage{graphicx}

\makeatletter

\newcommand*\LyXThinSpace{\,\hspace{0pt}}
\pdfpageheight\paperheight
\pdfpagewidth\paperwidth

\makeatother

\usepackage{babel}
\begin{document}
\title{Finite-time quantum measurement cooling beyond the Carnot limit}
\author{Tong Fu$^{1}$}
\author{Jianying Du$^{1}$}
\author{Jingyi Chen$^{1}$}
\author{Jincan Chen$^{1}$}
\author{Chikako Uchiyama$^{2}$}
\author{Shanhe Su$^{1}$}
\email{sushanhe@xmu.edu.cn}

\address{$^{1}$Department of Physics, Xiamen University, Xiamen, 361005, People's
Republic of China}
\address{$^{2}$Faculty of Engineering, University of Yamanashi, Kofu, Yamanashi
400-8511, Japan and National Institute of Informatics, Chiyoda, Tokyo
101-8430, Japan}
\begin{abstract}
We proposed the finite-time cycle model of a measurement-based quantum
cooler, where invasive measurement provides the power to drive the
cooling cycle. Such a cooler may be regarded as an alternative thought
experiment of Mawell\textquoteright s demon. The measurement-feedback
information is capable of moving heat from the cold to hot bath without
any work input and even making the maximum coefficient of performance
larger than the Carnot limit. The causes that this seemingly paradoxical
result does not violate the laws of thermodynamics can be clearly
explained through the derivation of a generalized Clausius inequality
including the mutual information.
\end{abstract}
\maketitle
For a classical system exchanging heat with two external baths and
undergoing a thermodynamic cycle, Clausius's inequality reads $-\left\langle Q_{h}\right\rangle /T_{h}-\left\langle Q_{c}\right\rangle /T_{c}\geq0$,
where $\left\langle Q_{\alpha}\right\rangle \left(\alpha=h,c\right)$
represents the amount of heat flowing into the system from bath $\alpha$
with temperature $T_{\alpha}$ \citep{Callen-book}. The inequality
indicates that it is impossible to construct a self-acting machine,
unaided by any external agency, to transfer heat from a cool bath
to a hot bath. Meanwhile, all thermal machines between two heat baths
are less efficient than the Carnot cycle. 

Nevertheless, quantum thermodynamics points to new amazing discoveries.
Nonequilibrium thermal reservoirs, including quantum coherent \citep{Scully2003,Quan2006}
and squeezed thermal reservoir \citep{Ro=0000DFnagel2014,Klaers2017,Huang2012,Manzano2016,Zhang2020},
are expected to be the resources for a quantum machine beyond the
thermodynamic bound of its standard counterpart. Niedenzu et al. revealed
an efficiency bound for the quantum engine with a non-thermal bath
by deriving a tight inequality between the entropy change of the system
and the energy exchanged with the bath \citep{Niedenzu2018,Niedenzu2018njp}.
Watanabe revealed that quantum statistics yields the enhancement of
work of thermal machine through many cycles and multiple work resources
\citep{Watanabe2020,Watanabe2017}. Shirai et al. considered the non-Markovian
effect in a quantum Otto engine and stated that Carnot's theorem is
consistent with a definition of work including the energy of the system-reservoir
interaction \citep{Shirai2021}. Micadei et al. experimentally demonstrated
that quantum correlation between the qubits allows the reversal of
heat flow \citep{Micadei2019}. For a time-dependent system in a non-adiabatic
quantum evolution, both Brandner and Su found that quantum coherence
plays an important role in heat and work \citep{Brandner2017,Su2018}.
The concept of fluctuating efficiency has been introduced, because
quantum-scale engines are subjected to thermal and quantum fluctuations
\citep{Verley2014,Denzler2021}. Pietzonka and Seifert showed that
power fluctuations determine the bounds on the power and efficiency
of a class of quantum machines through a universal trade-off relation
among the power, efficiency, and fluctuation \citep{Pietzonka2018,Koyuk2018}.

More interestingly, quantum measurement allows us to create a variety
of ingenious energy conversion processes. There exist two standard
ways of designing thermal machines with quantum measurement. The first
case is characterized the noninvasive measurement, which does not
alter the internal energy for the measured system because of the fact
that the measurement basis corresponds to the energy eigenbasis of
the Hamiltonian $H$ of the system \citep{Braginsky-book,Sun1993,Maruyama2009}.
In general, the measurement acquires information about the state of
the working substance, followed by a feedback control evolution depending
on the measurement outcome. Based on the thought experiments of Maxwell\textquoteright s
demon \citep{Maxwell-book}, Dong et al. translated Szilard\textquoteright s
classical engine \citep{Szilard1929,Zurek-book} into a quantum version
by using the noninvasive interaction between the measuring apparatus
and the system \citep{Dong2011}. Quan et al. designed a Maxwell\textquoteright s
demon assisted thermodynamic cycle, where the measurement is implemented
through a controlled-NOT gate operation and does not increase the
entropy \citep{Quan2006-1}. Manzano et al. assessed the roles of
classical and quantum correlations in the optimal work extraction
from a bipartite system by performing a measurement on the ancilla
unit \citep{Manzano2018}. Employing an effective nonselective projective
measurement in the energy basis of the $^{13}C$ nuclear spin, Camati
et al. provided an experimental evidence of the trade-off between
the information and the entropy production \citep{Camati2016}. Experimental
observations of the role of the noninvasive measurement in the work
extraction and the fluctuation have been conducted in superconducting
quantum circuits as well \citep{Koski2014,Masuyama2018}. The second
case is characterized by the invasive measurement, where the internal
energy of the observed system is changed if the measurement basis
is not identical to the eigenstate of $H$. Jacobs showed that this
additional energetic difference leads to a tight version of bound
on the extractable work from a system initially in thermal equilibrium
\citep{Jacobs2009}. Brandner et al. obtained a natural definition
for the efficiency of information to work conversion by extending
this bound to feedback-driven quantum engines running periodically
\citep{Brandner2015}. Elouard et al. proposed efficient quantum measurement
engines where work is directly extracted from the measurement channel
instead of a heat bath \citep{Elouard2017,Elouard2017-1,Elouard2018}.
The average work and efficiency of four-stroke quantum engines alternately
interacting with a measurement apparatus and a single heat bath have
been revealed \citep{Yi2017,Ding2018}. Measurement-driven machines
were extended to composite working substances. For a two-stroke two-qubit
cooler, the singlet-triplet basis maximizes the energy extraction
\citep{Buffoni2019}. An alternate scheme introduced a two-qubit engine
powered by entanglement and local measurements \citep{Bresque2021}.
Quantum discord has been related to the quantum feedback cooling by
controlling the state of the system through the measurement on ancillas
\citep{Liuzzo-Scorpo2016}. 

Note that the invasive measurement changes not only the internal energy
but also the entropy of the system. Most quantum thermodynamic cycles,
such as Carnot and Otto cycles, consider connecting cooling and heating
processes via adiabatic processes. In this Letter, we adopt a different
approach by replacing the adiabatic processes with the invasive measurement.
The energy offered by the measurement process will be regarded as
the work for pumping heat from a cold to hot bath. It would be interested
to know if this new configuration has a stable limit cycle. By including
the information gain in the measurement, it will be revealed whether
a generalized Clausius inequality leads to an unusual bound on the
performance for this cycle. Particularly, attention will be paid to
create a measurement assisted cooler with a coefficient of performance
(COP) larger than the Carnot limit even for finite-time scale. 

\textbf{Results.} For a quantum system alternately contacting with
thermal baths at temperatures $T_{h}$ and $T_{c}$ as shown in Fig.
1(a), the measurement-based feedback control results in the inequality

\begin{equation}
\left\langle \sigma\right\rangle =\left\langle I\right\rangle -\frac{\left\langle Q_{h}\right\rangle }{T_{h}}-\frac{\left\langle Q_{c}\right\rangle }{T_{c}}\geqslant0\text{,}\label{eq:in}
\end{equation}
where $\left\langle \sigma\right\rangle $ is the total amount of
entropy production, and $\left\langle I\right\rangle $ denotes the
mutual information associated with the measured system that has been
obtained by measurement. $\left\langle S\right\rangle =-\frac{\left\langle Q_{h}\right\rangle }{T_{h}}-\frac{\left\langle Q_{c}\right\rangle }{T_{c}}$
is interpreted as the average entropy change of the two baths. When
$\left\langle Q_{c}\right\rangle =-\left\langle Q_{h}\right\rangle >0$,
the information offers the possibility of cooling without needing
an energy input.

The bound on the COP $\varepsilon$ of a cooler could be established
by rearranging the inequality (\ref{eq:in}), i.e.,

\begin{equation}
\frac{\varepsilon}{\varepsilon_{C}}\leq1+\frac{T_{h}\left\langle I\right\rangle }{\left\langle W\right\rangle }\label{eq:cop}
\end{equation}
where $\varepsilon_{C}=T_{c}/\left(T_{h}-T_{c}\right)$ is the reversible
COP of a Carnot cycle, and the cooler is driven by energy $\left\langle W\right\rangle =-\left\langle Q_{h}\right\rangle -\left\langle Q_{c}\right\rangle \left(\left\langle Q_{h}\right\rangle \neq-\left\langle Q_{c}\right\rangle \right)$
provided by the invasive measurement. Equation (\ref{eq:cop}) gives
rise to a new limit of the COP that incorporates the information $\left\langle I\right\rangle $
. A main consequence of this new bound is that a quantum measurement
cooler may have a performance beyond the Carnot limit. If $\left\langle I\right\rangle =0$,
Eq. (\ref{eq:in}) reduces to the traditional Clausius inequality
and $\varepsilon$ is then bounded by the Carnot COP. 

\begin{figure}
\begin{centering}
\includegraphics[scale=0.3]{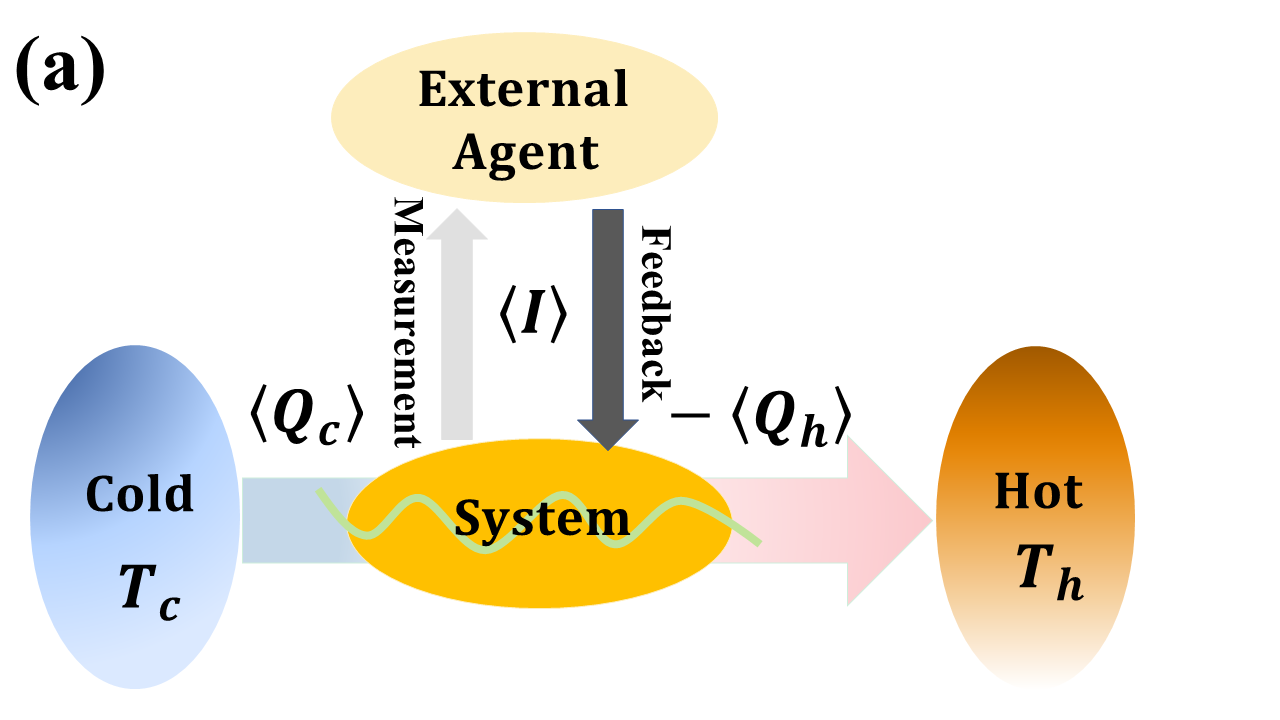}
\par\end{centering}
\centering{}\includegraphics[scale=0.25]{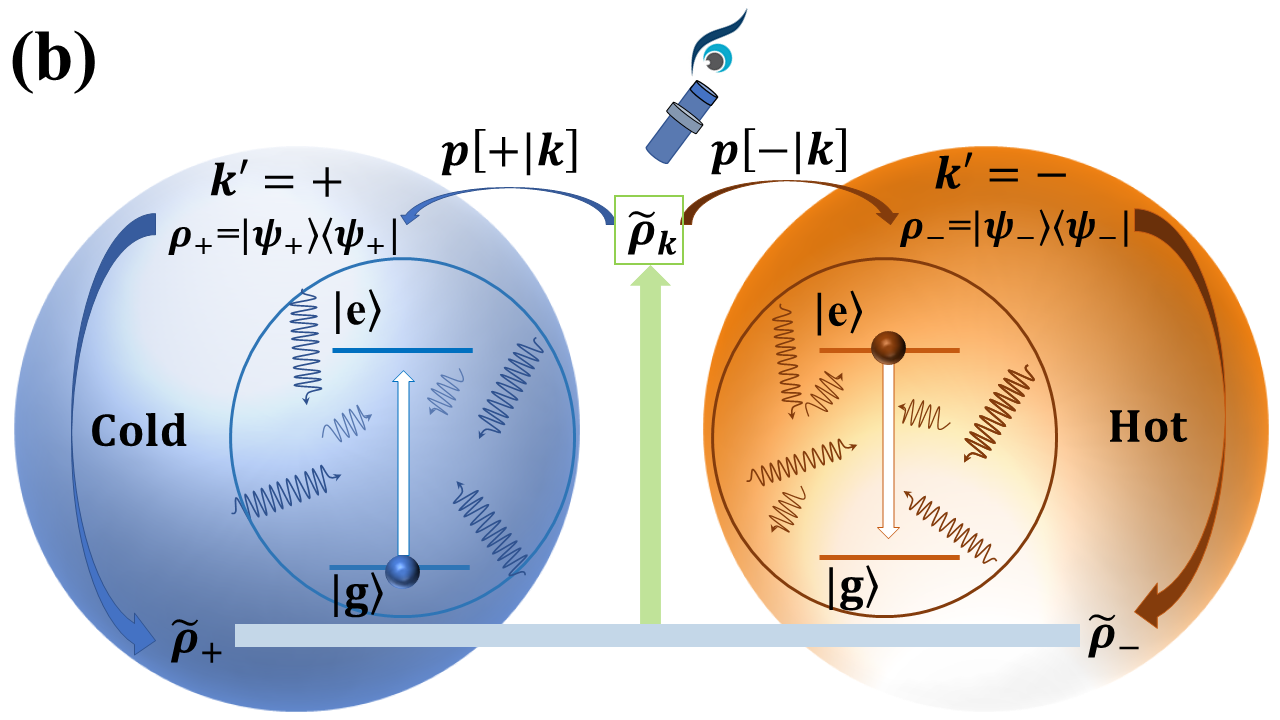}\caption{(a) The schematic of the quantum measurement cooler. (b) The control
protocol. }
\end{figure}

\textbf{Model.} As an application of this concept, we consider a two-level
system as the working substance. The Hamiltonian is assumed as the
form

\begin{equation}
H_{S}=\frac{\hbar\omega}{2}\sigma_{z},
\end{equation}
where $\hbar\omega>0$ is the energy spacing between the exited state
$|e\rangle$ and the ground state $|g\rangle$, and the Pauli operator
$\sigma_{z}=|e\rangle\langle e|-|g\rangle\langle g|$. The control
protocol of the cooling cycle is illustrated in Fig. 1(b). The two-level
system is prepared in an arbitrary initial state $\rho_{\mathrm{0}}$.
In the first stroke, the system interacts with a measurement apparatus,
projecting the system onto the basis $\left\{ \left|\psi_{+}\right\rangle ,\left|\psi_{-}\right\rangle \right\} $.
If the state vector after measurement is $\left|\psi_{k}\right\rangle \left(k=+,-\right)$,
the state of the system is updated to $\rho_{k}=\pi_{k}\rho_{0}\pi_{k}^{\dagger}/Tr\left\{ \pi_{k}\rho_{0}\pi_{k}^{\dagger}\right\} =\left|\psi_{k}\right\rangle \left\langle \psi_{k}\right|$
\citep{Elouard2018,Jordan2020,Breuer-book}, where $\pi_{k}=\left|\psi_{k}\right\rangle \left\langle \psi_{k}\right|$
represents the orthogonal projector associated with the measurement
basis. 

In the second stroke, an isochoric process is executed through the
feedback control. The Lindblad master equation will be introduced
to treat the time evolution of the system interacting with a thermal
bath \citep{Breuer-book,Breuer2016}. The feedback operator $\mathcal{V}_{k}$
is performed with probability $p_{k}^{0}=\left\langle \psi_{k}\left|\rho_{\mathrm{0}}\right|\psi_{k}\right\rangle $,
leading to the evolved density matrix $\tilde{\rho}_{k}=\mathcal{V}_{k}\left[\rho_{k}\right]$
under a trace-preserving completely positive map \citep{Nielsen-book}.
The first case is characterized by the postmeasurement state vector
$\left|\psi_{+}\right\rangle $, where the system is put in contact
with a cold bath. For the outcome with the state vector $\left|\psi_{-}\right\rangle $,
the system interacts with a dissipative hot bath. Let $\tau_{c}$
$\left(\tau_{h}\right)$ be the time interval during which the system
is brought in thermal contact with the cold (hot) bath. The density
matrix at the end of the feedback control serves as the initial state
of the succeeding cycle. 

The two thermodynamic processes are carried out alternately until
the measurement outcomes reach a steady distribution. For the two
state system, the constant probability $p_{+}$ of the state vector
$\left|\psi_{+}\right\rangle $ has the form (Supplementary I)

\begin{equation}
p_{+}=\frac{q[+|-]}{q[+|-]-q[+|+]+1},\label{eq:cp+}
\end{equation}
where $q\left[k^{\prime}|k\right]$ represents the conditional probability
of the measurement outcomes $\left|\psi_{k^{\prime}}\right\rangle $
in one cycle and $\left|\psi_{k}\right\rangle $ in the previous cycle,
i.e., 
\begin{equation}
q\left[k^{\prime}|k\right]=\left\langle \psi_{k^{\prime}}\left|\tilde{\rho}_{k}\right|\psi_{k^{\prime}}\right\rangle =\left\langle \psi_{k^{\prime}}\left|\mathcal{V}_{k}\left[\left|\psi_{k}\right\rangle \left\langle \psi_{k}\right|\right]\right|\psi_{k^{\prime}}\right\rangle .
\end{equation}

By considering $p\left(k^{\prime},k\right)\equiv q\left[k^{\prime}|k\right]p_{k}$
as the probability of two consecutive measurement outcomes $\left|\psi_{k}\right\rangle $
and $\left|\psi_{k^{\prime}}\right\rangle $ and the sum rule $\sum_{k^{\prime}}p\left[k^{\prime}\mid k\right]=1$,
the average work performed by the measurement
\begin{equation}
\left\langle W\right\rangle =\sum_{k}p_{k}\left[\left\langle \psi_{k}|H_{S}|\psi_{k}\right\rangle -Tr\left\{ H_{S}\tilde{\rho}_{k}\right\} \right].\label{eq:w}
\end{equation}
The average entropy change of the system per cycle associated with
the measurement reads

\begin{equation}
\left\langle \Delta\mathrm{S}^{\mathrm{m}}\right\rangle =k_{\mathrm{B}}\sum_{k}p_{k}Tr\left\{ \tilde{\rho}_{k}\ln\tilde{\rho}_{k}\right\} .\label{eq:sm-1}
\end{equation}

After the measurement, the system has the probability $p_{+}$ to
contact with the cold bath. The average heat extracted from the cold
bath 

\begin{equation}
\left\langle Q_{c}\right\rangle =p_{+}\left(Tr\left\{ H_{S}\tilde{\rho}_{+}\right\} -Tr\left\{ H_{S}\rho_{+}\right\} \right).\label{eq:qc}
\end{equation}
Similarly, the system gets in touch with the hot bath at the probability
$p_{-}$, resulting in the average heat extracted from the hot bath
as 

\begin{equation}
\left\langle Q_{h}\right\rangle =p_{-}\left(Tr\left\{ H_{S}\tilde{\rho}_{-}\right\} -Tr\left\{ H_{S}\rho_{-}\right\} \right).\label{eq:qh}
\end{equation}
Note that $\left\langle Q_{c}\right\rangle +\left\langle Q_{h}\right\rangle +\left\langle W\right\rangle =0$
satisfying the first law of thermodynamics. The more detailed calculations
of Eqs. (\ref{eq:w})-(\ref{eq:qh}) are presented in Supplementary
II.

The entropy productions of both types of control evolution are always
positive, resulting in the average entropy production (Supplementary
III)

\begin{align}
\left\langle \sigma\right\rangle  & =-k_{\mathrm{B}}\sum_{k^{\prime}}p_{k^{\prime}}Tr\left\{ \tilde{\rho}_{k^{\prime}}\ln\tilde{\rho}_{k^{\prime}}\right\} +k_{\mathrm{B}}\sum_{k}p_{k}Tr\left\{ \rho_{k}\ln\rho_{k}\right\} \nonumber \\
 & -\frac{\left\langle Q_{h}\right\rangle }{T_{h}}-\frac{\left\langle Q_{c}\right\rangle }{T_{c}}\geqslant0\label{eq:INE}
\end{align}
The sum of the first two terms can be interpreted as the average mutual
information $\left\langle I\right\rangle $ describing the information
about the measured system that has been obtained by measurement \citep{Camati2016,Sagawa2008}.
Therefore, the quantum cooler satisfies the inequality Eq. (\ref{eq:in}).
Once the periodically measurement-driven refrigerator is established,
information may become a source of energy to move the heat from the
cold to hot bath. 

\textbf{Discussion.} For the first stroke, we choose the measurement
bases $\left|\psi_{k}\right\rangle $ associated with measurement
operator $\pi_{k}$ as $\left|\psi_{+}\right\rangle =\cos\frac{\vartheta}{2}|e\rangle+e^{i\varphi}\sin\frac{\vartheta}{2}|g\rangle$
and $\left|\psi_{-}\right\rangle =e^{-i\varphi}\sin\frac{\vartheta}{2}|e\rangle-\cos\frac{\vartheta}{2}|g\rangle$,
where $\vartheta$ and $\varphi$ represent, respectively, the colatitude
with respect to the $\mathrm{z}$-axis and the longitude with respect
to the $x$-axis in the Bloch sphere representation. In the feedback
control process, a two-level system that is damped by interacting
with the radiation field in thermal equilibrium at temperature $T_{\alpha}$
is adopted. The Hamiltonian $H_{\mathrm{\alpha}}$ of bath $\alpha$
and the interaction Hamiltonian $H_{S-\alpha}$ between the system
and the bath are given in Supplementary V. With the Born-Markov approximation,
one obtains the evolution of the system under the feedback control
$\mathcal{V}_{+}$ in the Kraus representation (Supplementary V),
where we have defined the Plank distribution of the cold bath $n_{c}=1/\left\{ \exp\left[\hbar\omega/\left(k_{B}T_{c}\right)\right]-1\right\} $,
the coupling parameter $\gamma_{c}$, and $\Gamma_{c}=\gamma_{c}(2n_{c}+1)$.
The evolution of system under the feedback control $\tilde{\rho}_{-}=\mathcal{V}_{-}\left[\rho_{-}\right]$
is obtained by replacing the subscript $c$ with $h$.

\begin{figure}
\begin{centering}
\includegraphics[scale=0.3]{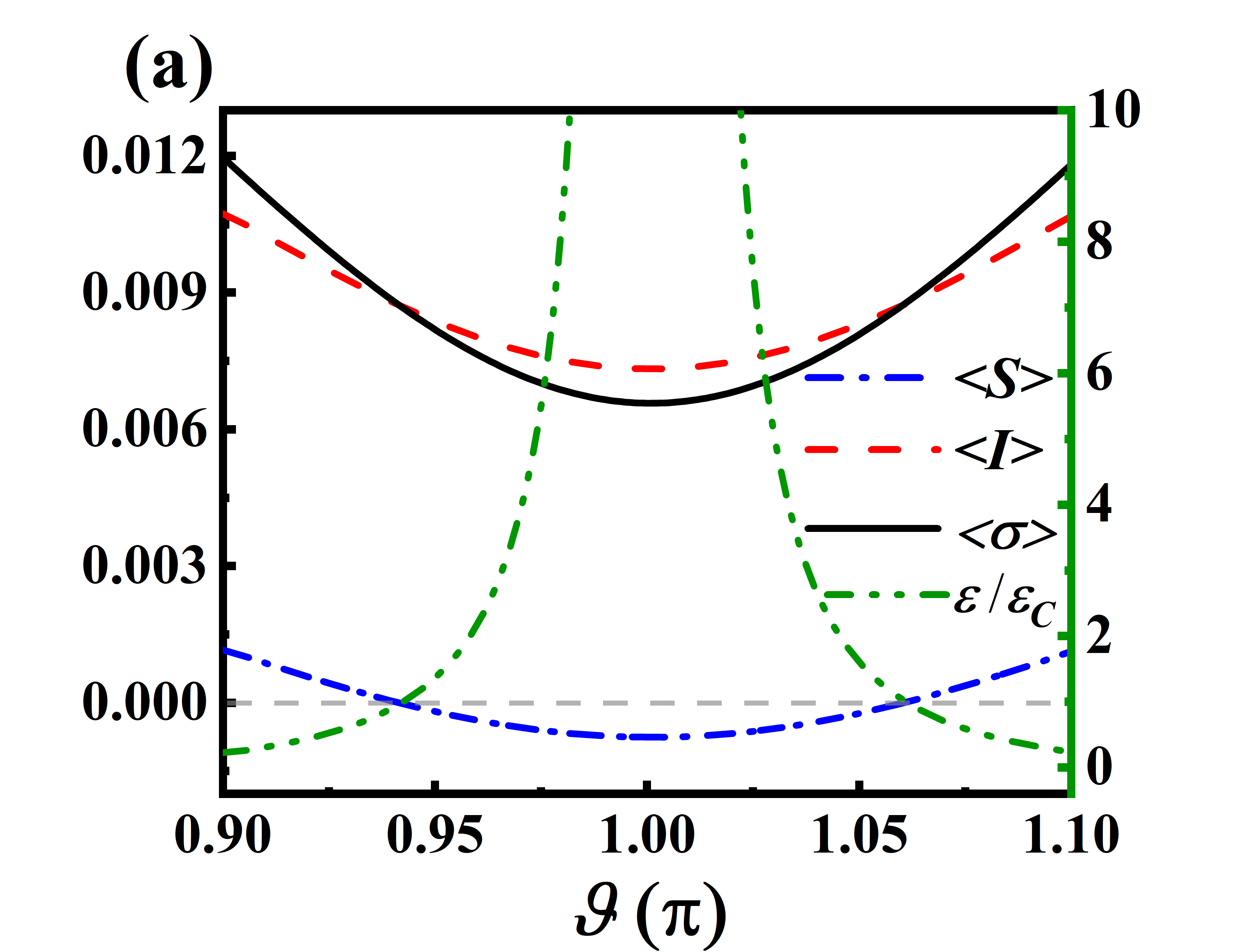}
\par\end{centering}
\centering{}\includegraphics[scale=0.3]{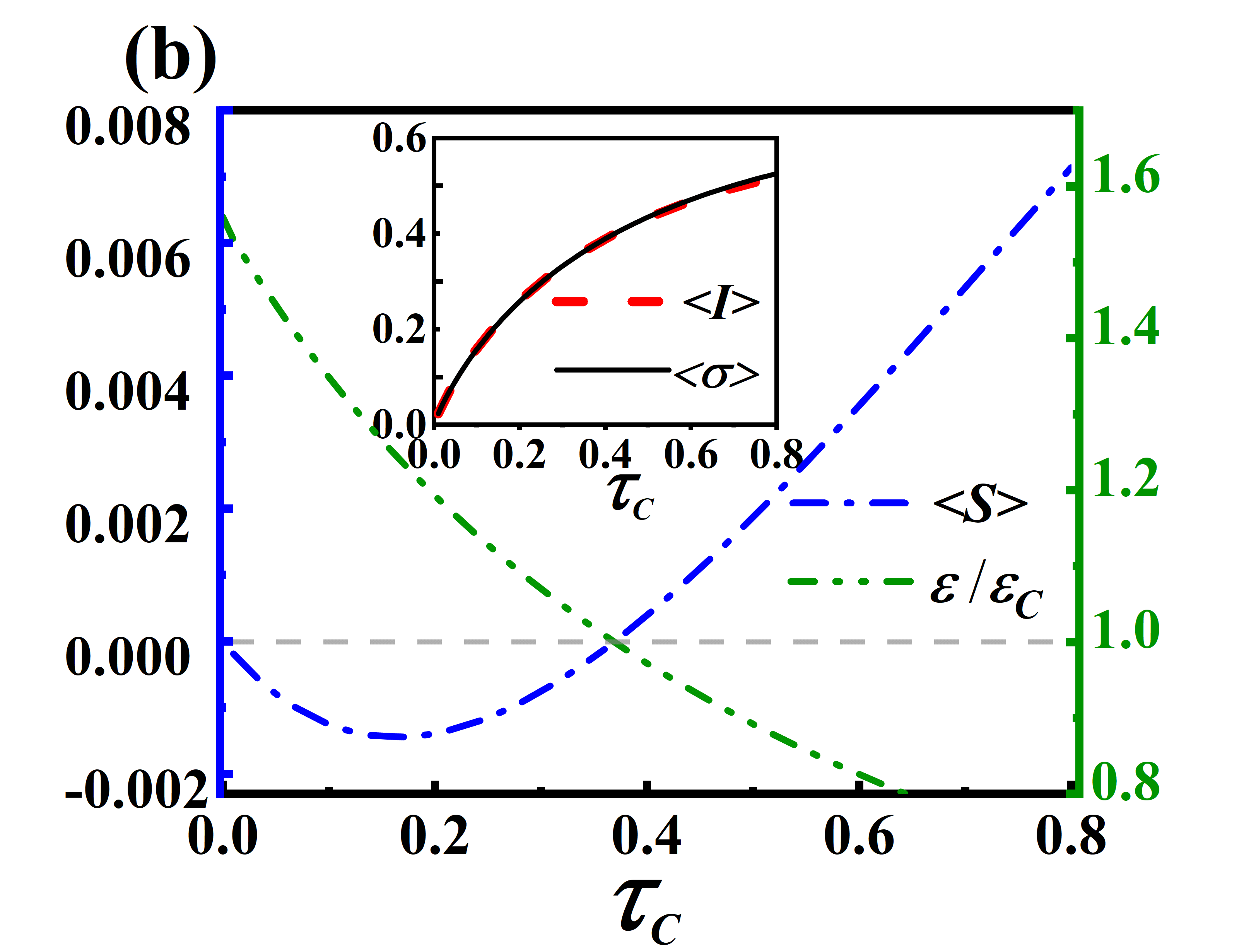}\caption{(a) The total amount of entropy production $\left\langle \sigma\right\rangle $,
mutual information $\left\langle I\right\rangle $, average entropy
change $\left\langle S\right\rangle $ of the hot and cold baths,
and dimensionless COP $\varepsilon\text{/\ensuremath{\varepsilon_{C}}}$
varying with the colatitude $\vartheta$, where $\tau_{c}=0.5$, the
left vertical axis shows the values for $\left\langle \sigma\right\rangle $,
$\left\langle I\right\rangle $, and $\left\langle S\right\rangle $,
and the corresponding scales of $\varepsilon\text{/\ensuremath{\varepsilon_{C}}}$
is on the right vertical axis. Note that the performance of the cycle
is independent on the longitude $\varphi$. (b) The average entropy
change $\left\langle S\right\rangle $ of the hot and cold baths varying
with the time interval $\tau_{c}$, where $\vartheta=0.98\pi$, the
left vertical axis shows the value for $\left\langle S\right\rangle $,
and the corresponding scales of $\varepsilon\text{/\ensuremath{\varepsilon_{C}}}$
is on the right vertical axis. The inserted figure presents $\left\langle \sigma\right\rangle $
and $\left\langle I\right\rangle $ varying with $\tau_{c}$. The
rest parameters $\omega=0.5$,\textcolor{black}{{} $T_{h}=0.2,T_{c}=0.1,$
$\gamma_{h}=\gamma_{c}=0.01$, $\tau_{h}=1$, and $\varphi=\frac{\pi}{4}$,
where $\hbar=k_{B}=1$.}}
\end{figure}

Fig.2 (a) indicates that the average entropy change of the baths $\left\langle S\right\rangle $
(blue dash-dotted line) may be less than zero and the COP of the cooler
is larger than the Carnot COP, i.e., $\varepsilon\text{/\ensuremath{\varepsilon_{C}}}\geq1$
(green dash-double-dotted line). A consideration of the mutual information
$\left\langle I\right\rangle $ (red dash line) guarantees that the
entropy production $\left\langle \sigma\right\rangle \geq0$ (black
solid line). In the case of $\left\langle S\right\rangle =0$, the
definition of $\left\langle S\right\rangle $ results in $\varepsilon=\ensuremath{\varepsilon_{C}}$
(the crosspoints between the blue dash-dotted line and the green dash-double-dotted
line). When the colatitude $\vartheta=\pi$, \textcolor{black}{the
measurement bases reduce to the eigenstates of $H$, i.e., $\left|\psi_{+}\right\rangle \propto|g\rangle$
and $\left|\psi_{-}\right\rangle \propto|e\rangle$}. The measurement
does not change the internal energy of the system. However, heat flows
from the cold to hot bath without any work input, i. e., $\left\langle W\right\rangle =0$
{[}Eq. (\ref{eq:w}){]}, leading to an infinitely large COP $\varepsilon\rightarrow\infty$.
The measurement is likely to prepare the system in the ground state
$|g\rangle$ before contacting with the cold bath and in the excited
state before getting in touch with the hot bath. Therefore, energy
is always extracted from the cold bath and released to the hot bath.
\textcolor{black}{In fact,} the cycle acts as a cooler driven by purely
information, representing an alternative thought experiment of Mawell's
demon.\textcolor{red}{{} }

In Fig. 2(b), $\left\langle S\right\rangle $ (blue dash-dotted curve)
is initially a decreasing function of the time interval $\tau_{c}$.
After reaching a minimum value, $\left\langle S\right\rangle $ will
increase with $\tau_{c}$. By implementing an external agent with
measurement and feedback, it is no doubt that $\varepsilon$ goes
beyond the Carnot limit $\varepsilon_{C}$ in the region of $\left\langle S\right\rangle <0$.
The mutual information $\left\langle I\right\rangle $ (red dash line)
ensures $\left\langle \sigma\right\rangle \geq0$ (black solid line)
in this area. Similar to Fig. 2(a), the point of $\left\langle S\right\rangle =0$
coincides with that of $\varepsilon=\varepsilon_{C}$. 

When the times $\tau_{c}\rightarrow\infty$ and $\tau_{h}\rightarrow\infty$,
or $\gamma_{c}\rightarrow\infty$ and $\gamma_{h}\rightarrow\infty$,
thermal equilibrium states under the feedback controls $\mathcal{V}_{+}$
and $\mathcal{V}_{-}$ are achieved. The conditional probabilities
$q[+|+]$ and $q[+|-]$ are, respectively, simplified as $q[+|+]=\left[(1+2n_{c})-\cos\vartheta\right]/\left[2(1+2n_{c})\right]$
and $q[+|-]=\left(1+2n_{h}-\cos\vartheta\right)/\left[2(1+2n_{h})\right]$.
By using Eqs. (\ref{eq:cp+}), (\ref{eq:qc}), and (\ref{eq:qh}),
the inequality $\left\langle S\right\rangle \leq0$ can be equivalently
expressed as

\begin{equation}
-\frac{\left\langle Q_{c}\right\rangle }{\left\langle Q_{h}\right\rangle }=\frac{\gamma_{h}\left(n_{h}+\sin^{2}\frac{\vartheta}{2}\right)\left(\cos^{2}\frac{\vartheta}{2}+n_{c}\cos\vartheta\right)}{\gamma_{c}\left(n_{c}+\cos^{2}\frac{\vartheta}{2}\right)\left(-\sin^{2}\frac{\vartheta}{2}+n_{h}\cos\vartheta\right)}\geq\frac{T_{c}}{T_{h}}.\label{eq:inq}
\end{equation}
As the colatitude $\vartheta$ makes the ratio between the heat extracted
from the cold bath $\left\langle Q_{c}\right\rangle $ and that extracted
from the hot bath $\left\langle Q_{h}\right\rangle $ meet the condition
given by Eq. (\ref{eq:inq}), the COP $\varepsilon$ will be greater
than the Carnot limit $\varepsilon_{C}$. Particularly, an algebra
calculation shows that $\left\langle Q_{c}\right\rangle /\left\langle Q_{h}\right\rangle =-1$
at $\vartheta=\pi$ and $\gamma_{h}=\gamma_{c}$, leading to $\varepsilon\rightarrow\infty$. 

In conclusions, a general bound on the COP of quantum measurement
coolers is presented. The mutual information obtained by measurement
has the potential to improve the cooling performance. When the average
entropy changes of the two baths are negative, the COP of the two-stoke
cycle consisting of measurement and feedback surpasses the Carnot
limit. The mutual information ensures that the entropy production
is always positive, satisfying the second law of thermodynamics. In
the case of a measurement basis corresponding to the energy eigenstates
of the system, the feedback control makes heat flow from the cold
to hot bath without any work input. The results derived from the present
finite time model offer a broad configuration in the application of
the fuel energy from quantum measurement and provide more instructive
information to the development of experiments than those obtained
by traditional time-independent models. 

\textbf{Acknowledgments}

This work has been supported by the National Natural Science Foundation
(Grants No. 12075197 and No. 11805159) and the Fundamental Research
Fund for the Central Universities (No. 20720210024).

\end{document}